\documentclass[a4paper]{jpconf}
\usepackage{graphicx}
\usepackage{here}
\usepackage{subcaption}
\begin{document}
\title{Charged Higgs boson production via $pp \to H^\pm bj$ at the LHC}

\author{Rachid Benbrik$^1$, Mohamed Krab$^2$ and Mohamed Ouchemhou$^1$}

\address{$^1$ Laboratory of Fundamental and Applied Physics, Polydisciplinary Faculty of Safi, Sidi Bouzid, BP 4162, Safi, Morocco.}
\address{$^2$ Research Laboratory in Physics and Engineering Sciences, Modern and Applied Physics Team, Polidisciplinary Faculty, Beni Mellal, 23000, Morocco.}

\ead{r.benbrik@uca.ma, mohamed.krab@usms.ac.ma, mohamed.ouchemhou@ced.uca.ac.ma}

\begin{abstract}
	The searches for charged Higgs bosons can be used to probe new physics at the LHC. In the current study, we concentrate on the associated production of the charged Higgs boson with the bottom quark and the jet in Two-Higgs Doublet Model (2HDM) type-I as promising mode for a light $H^\pm$, i.e. $m_{H^\pm}<m_t$. For this we consider the two situations where $h$ or $H$ is the SM-type Higgs boson discovered with a mass close to 125 GeV and investigate its bosonic decays, namely $H^{\pm} \to W^{\pm} h$ and/or $H^{\pm} \to W^{\pm} A$. The potential signals at the LHC are then discussed, under both theoretical and experimental constraints. Over a substantial region of the 2HDM type-I parameter space, a promising alternative signals that could be used for the discovery of $H^{\pm}$ states at the LHC are identified as $qbW+2b/2\tau/2\gamma$.

\end{abstract}

\section{Introduction}
In 2012, the discovery of the Higgs particle at the Large Hadron Collider (LHC) \cite{ATLAS:2012yve,CMS:2012qbp} has provided many opportunities to track down new physics beyond the Standard Model. Among the candidates of special and promising interest is the charged Higgs Boson, because it is not part of the Standard Model (SM), so its discovery would constitute clear scientific evidence for new physics. The most simple extension of the SM that leads to charged Higgs bosons is the Two-Higgs Doublet Model (2HDM) with a softly broken $Z_2$ symmetry. The $Z_2$ symmetry is incorporated to prevent flavor-changing neutral currents (FCNCs) at the tree-level, and it yields four distinct models, referred to here as type I, type II, type X and type Y respectively. Both direct and indirect constraints on the charged Higgs boson, the former has been provided by the LEP experiment requiring its mass to be greater than $80$ GeV at $95 \%$ CL, and the latter is model dependent, and may be derived from B meson decays, and in particular those from $B \to X_s\gamma$ \cite{Haller:2018nnx}, a charged Higgs boson with a mass below 650 GeV almost entirely independent of $\tan\beta$ in the type II and type Y models has been excluded. However, the type I and X models are still allowing for light charged Higgs bosons below than top quark mass.

The charged Higgs phenomenology has been extensively examined at collider experiments (LHC), it's most common production modes includes: the light charged higgs production $pp \to t\bar{t}$ process via the top(anti-top) decay $t \to bH^+(\bar{t}\to \bar{b}H^-)$ mode with $m_{H^{\pm}} \leq m_t - m_b$, competing with the SM decay of $\bar{t}\to W^- \bar{b}$ that's provide the largest production rate for such light
charged Higgs bosons when it's kinematically allowed via the process $pp \to t\bar{t}\to b\bar{b}H^- W^+ + \rm{C.C.}$, in addition, the light charged Higgs can be produced via the following modes:  production in association with a top and a bottom quark can be calculated either in the four-flavor scheme via the process $gg \to t\bar{b}H^+$ or in the five flavor scheme via the process $g\bar{b} \to \bar{t}H^+$ \cite{Alwall:2004xw},  production in association with a charged gauge boson via $b\bar{b}\to H^\pm W^\mp$ at tree level and $gg\to H^\pm W^\mp$ at loop level \cite{Moretti:1998xq,BarrientosBendezu:1998gd}, production associated with a bottom quark and a light quark $qb\to q^{'}H^+ b$ \cite{Arhrib:2015gra}, resonant production via the process $c\bar{s}, c\bar{b} \to H^+$ \cite{Hernandez-Sanchez:2012vxa,Hernandez-Sanchez:2020vax}, associate production with a neutral Higgs via the process $q\bar{q} \to H^\pm \Phi~(\Phi= h, H,A)$ \cite{Enberg:2018nfv}, and the pair production through $q\bar{q} \to H^+ H^-$ annihilation process or gluon fusion $gg \to H^+ H^-$ \cite{Moretti:2001pp,Moretti:2003px}. The goal of this contribution is to look at the generation of singly-charged Higgs bosons in conjunction with a bottom quark and a jet $j$ using the subprocess $pp(qb) \to  H^+ b j$ in the 2HDM type-I.

This work is organized as follows, we briefly introduce the 2HDM theoretical framework in Section \ref{section1}, followed by the scan parameters and constraints applied in Section \ref{section2},  the results and discussion will take place in Section \ref{section4}. Finally, we will come to our conclusion in Section \ref{section5}

\label{section1}
\section{The Two-Higgs Doublet Model}
With a softly broken $Z_2$ symmetry, the general 2HDM potential, which is renormalisable and invariant under $SU(2)_L \otimes U(1)_Y$, can be expressed as, 
\begin{eqnarray}
V_{\rm{2HDM}}(\Phi_1,\Phi_2)= & m_{11}^2(\Phi_1^+\Phi_1)+m_{22}^2(\Phi_2^+\Phi_2)-m_{12}^2(\Phi_1^+\Phi_2+h.c.)\nonumber\\
&+\lambda_1(\Phi_1^+\Phi_1)^2+\lambda_2(\Phi_2^+\Phi_2)^2+\lambda_3(\Phi_1^+\Phi_1)(\Phi_2^+\Phi_2)\nonumber\\
&+\lambda_4 (\Phi_1^+\Phi_2)(\Phi_2^+\Phi_1)+\frac{\lambda_5}{2}[(\Phi_1^+\Phi_2)^2+ \rm{h.c.}],\qquad \nonumber\\
\end{eqnarray}
where, $\Phi_i(i=1,2)$ are the two complex $SU(2)$ Higgs doublets, $m_{11}^2,m_{22}^2$ and $m_{12}^2$ are squared mass parameters, and $\lambda_{1-5}$ are dimensionless coupling parameters. The Vacuum Expectation Values (VEVs) of the Higgs doublet fields are $v_1$ and $v_2$, respectively, and satisfy $v_1^2 + v_2^2 = v^2 \approx (246\ \rm{GeV})^2$.  The $v_{1,2}$ of the two Higgs fields (which are fixed by the electroweak scale) can be substituted for the quartic couplings $\lambda_{1-5}$ by the four physical Higgs boson masses $m_h, m_H, m_A, m_{H^\pm}$  and the mixing angle $\alpha$ and $\beta$ using the minimization conditions of the potential, leaving us with the following seven free independent real parameters, 

\begin{equation}
m_h,\, m_H,\, m_A,\, m_{H^\pm},\, \sin(\beta-\alpha),\, \tan\beta,\, m_{12}^{2}.
\end{equation}
In the CP-odd and charged Higgs sectors, the angle $\beta$ is the rotation angle from the group eigenstates to the mass eigenstates. The comparable rotation angle for the CP-even Higgs sector is $\alpha$. At the tree level, (pseudo)scalars can be used to mediate FCNCs. The constraint of vanishing this FCNCs leads to impose a $Z_2$ symmetry, in which each type of fermion couples to just one of the Higgs doublets $\Phi_{1,2}$. The $Z_2$ charges can be assigned in four different ways, as a results, there are four different version of 2HDM \cite{Branco:2011iw}, known as type I, type II, type X or lepton-specific and type Y or flipped. In the type I model, the $Z_2$ symmetry is an accurate symmetry, and the $\Phi_1$ doublet imparts mass to all fermions. In the type II paradigm, the $\Phi_1$ doublet contributes mass to leptons and down quarks, while the $\Phi_2$ couples to up-type quarks. In the type X model, all quarks couple to the $\Phi_2$ doublet, while all charged leptons couple to the $\Phi_1$ doublet. Finally, in the type Y model, down-type quarks couple to the $\Phi_1$ doublet, whereas the remainder of the fermions couple to the $\Phi_2$ doublet. However, we only examine 2HDM type I in this work.

In the mass eigenstate basis, the Yukawa Lagrangian, which represents interactions between the Higgs and fermion sectors, is given by,

\begin{eqnarray}
- {\mathcal{L}}_{\rm Yukawa} = \sum_{f=u,d,l} \left(\frac{m_f}{v} \xi_f^h \bar{f} f h + 
\frac{m_f}{v}\xi_f^H \bar{f} f H 
- i \frac{m_f}{v} \xi_f^A \bar{f} \gamma_5 f A \right) + \nonumber \\
\left(\frac{V_{ud}}{\sqrt{2} v} \bar{u} (m_u \xi_u^A P_L +
m_d \xi_d^A P_R) d H^+ + \frac{ m_l \xi_l^A}{\sqrt{2} v} \bar{\nu}_L l_R H^+ + H.c. \right),
\label{Yukawa-1}
\end{eqnarray}
where the coefficients $\xi_{f}^{hi}$ are interpreted as the ratio of the Higgs boson couplings to the fermions with respect to the SM values, which are defined in the alignment limit for the type I in Table 1.

%=========================Tableau 
{\renewcommand{\arraystretch}{1.5} %donne la distance entre les lignes%
	{\setlength{\tabcolsep}{0.1cm} %donne la distance entre les collones%
\begin{table}[H]
	\caption{\label{coupling}Yukawa coupling coefficients of the neutral Higgs bosons $h$, $H$, $A$ to the up-quarks, down-quarks and the charged leptons $(u, d, l)$ in the 2HDM type I.}
	\begin{center}
		\begin{tabular}{cccccccccc}
			\br
            Type&$\xi_{u}^{h}$&$\xi_{d}^{h}$&$\xi_{l}^{h}$&$\xi_{u}^{H}$&$\xi_{d}^{H}$&$\xi_{l}^{H}$&$\xi_{u}^{A}$&$\xi_{d}^{A}$&$\xi_{l}^{A}$\\\hline
            I&$c_\alpha/s_\beta$&$c_\alpha/s_\beta$&$c_\alpha/s_\beta$&$s_\alpha/s_\beta$&$s_\alpha/s_\beta$&$s_\alpha/s_\beta$&$cot\beta$&$-cot\beta$&$-cot\beta$\\
			\br
		\end{tabular}
	\end{center}
\end{table}
\label{section2}
\section{Space parameter and constraints}
In order to investigate the phenomenology of the $H^\pm bj$ production mode in the 2HDM type I, we perform a random scan of the parameter space assuming one of the CP-even Higgs bosons $h$ or $H$ is the SM-like Higgs boson observed at the LHC with mass near $125$ GeV.

\begin{table}[H]
	\caption{\label{parameter-scan} 2HDM type I input parameters for both scenarios. All masses are in GeV.}
	\begin{center}
		\begin{tabular}{c c c c c c c c}
			\br
			parameter & $m_h$ & $m_H$ & $m_A$ &	$m_{H^\pm}$ & $\sin(\beta-\alpha)$ & $\tan\beta$ & $m_{12}^2 $ \\\hline
			$h$ scenario &  $125.09$ & $[126;\,200]$ & $[60;\,200]$ & $[80;\,170]$ & $[0.95;\,1]$ & $[2;\,15]$ & $[0;\,m_H^2\cos\beta\sin\beta]$\\\hline
			$H$ scenario & $[10;\,120]$ & $125.09$ & $[60;\,200]$ & $[80;\,170]$ & $[-1;\,1]$ & $[2;\,15]$ & $[0;\,m_h^2\cos\beta\sin\beta]$\\
			\br
		\end{tabular}
	\end{center}
\end{table}

Different conditions restrict the parameter space illustrated in Table \ref{parameter-scan}, using the 2HDMC public tool \cite{Eriksson:2009ws}. On the one hand, the model's correct high-energy behavior is ensured by unitarity, perturbativity, and vacuum stability. on the other hand, agreement with electroweak precision tests through the parameters $S$, $T$ and $U$ reduce the permissible mass splitting between heavy scalar fields. Apart from these model-structure-related constraints, the allowed parameter space is further constrained by using the codes HiggsSignals-2.6.0 \cite{Bechtle:2020uwn} and HiggsBounds-5.9.0 \cite{Bechtle:2020pkv} to enforce compatibility with the average LHC Higgs signal strength and the direct collider mass bounds on the heavy neutral and charged Higgs bosons, respectively. Finally, muon $(g-2)_\mu$ data and low-energy heavy flavor physics impose further indirect limitations on the ($m_{H^\pm},\ tan\beta$) plane. Using the tool SuperIso v4.1 \cite{Mahmoudi:2008tp}, we carefully included all of these limitations in our research. 

%============================= h-senario ===
\label{section3}
\section{Results and discussion}

In this work, we look for light charged Higgs boson traces in events involving a jet and a bottom quark, i.e. $pp \rightarrow qb \rightarrow H^\pm bj$. This light charged Higgs boson is motivated following the bosonic decays, $H^\pm \to W h$ and/or $H^\pm \to W A$, where $h$ and $A$ decay into a pair of bottom quarks, a pair of tau-lepton, or a pair of photons. Our interest here is the $qbW + b\bar{b}, \tau\tau$ and $\gamma\gamma$ final states in the $h$ and $H$ scenarios. Relevant LHC traces are summarized in Table \ref{traces}.
{\renewcommand{\arraystretch}{1.5} %donne la distance entre les lignes%
	{\setlength{\tabcolsep}{0.8cm} %donne la distance entre les collones%
\begin{table}[H]
	\caption{\label{traces} Production mechanisms and ultimate states of the charged Higgs boson. $\sigma_{H^\pm bj}$ stands for $\sigma(pp \rightarrow H^\pm b j)$ at $\sqrt{s} = 14$ TeV, and BR for branching ratio. $h_i~(i=1, 2)$ denotes $h_1=h$ in the $H$ scenario and $h_2=A$ in the $h$ scenario (see Table \ref{parameter-scan}). }
	\begin{center}
		\begin{tabular}{cc}
			\br
			\multicolumn{2}{c}{Higgs Production and Decay Process}\\\hline 
			$\sigma(pp \to qbW+b\bar{b})$ &$\sigma_{H^\pm bj}\times BR(H^\pm\to Wh_i)\times BR(hi\to b\bar{b})$\\\hline
		    $\sigma(pp \to qbW+\tau\tau)$ &$\sigma_{H^\pm bj}\times BR(H^\pm\to Wh_i)\times BR(hi\to \tau\tau)$\\\hline
            $\sigma(pp \to qbW+\gamma\gamma)$ &$\sigma_{H^\pm bj}\times BR(H^\pm\to Wh_i)\times BR(hi\to \gamma\gamma)$\\
			\br
		\end{tabular}
	\end{center}
\end{table}

\begin{figure}[H]
	\centering
	\begin{subfigure}[t]{.48\textwidth}\centering
		\includegraphics[height=5.8cm,width=7.8cm]{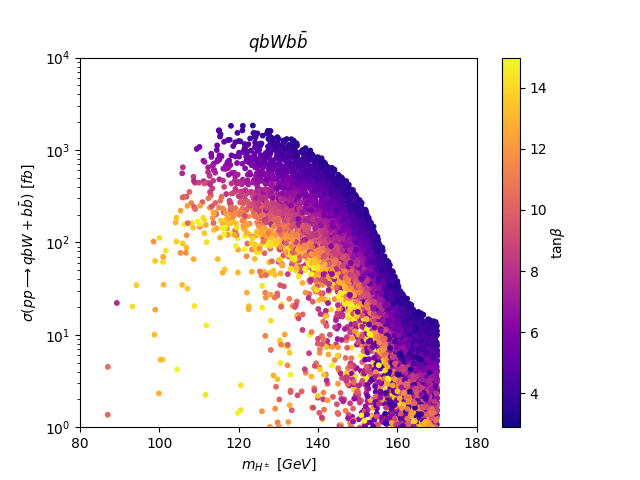}
		\caption{}
	\end{subfigure}
	\hfill
	\begin{subfigure}[t]{.48\textwidth}\centering
		\includegraphics[height=5.8cm,width=7.8cm]{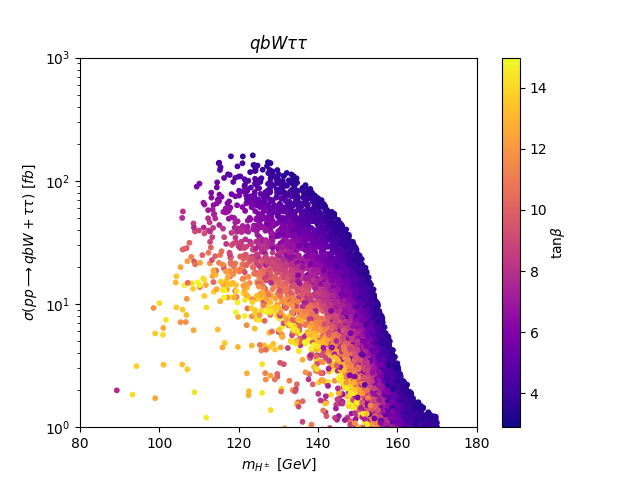}
		\caption{}
	\end{subfigure}
	\\[1cm]
	\begin{subfigure}[t]{.48\textwidth}\centering
		\includegraphics[height=5.8cm,width=7.8cm]{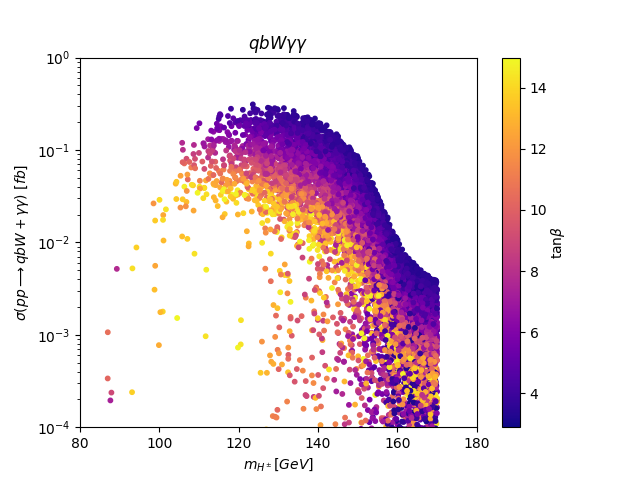}
		\caption{}
	\end{subfigure}
	\caption{\label{fig1} (a) $\sigma(pp\to qbW+\tau\tau)$, (b) $\sigma(pp\to qbW+\tau\tau)$ and (c) $\sigma(pp\to qbW+\gamma\gamma)$ showed against the charged Higgs mass in the $h$ scenario, with the color code indicates the parameter $\tan\beta$.}
\end{figure}
%%
%============================  H-senario ===

%
\begin{figure}[H]
	\centering
	\begin{subfigure}[t]{.48\textwidth}\centering
		\includegraphics[height=5.8cm,width=7.8cm]{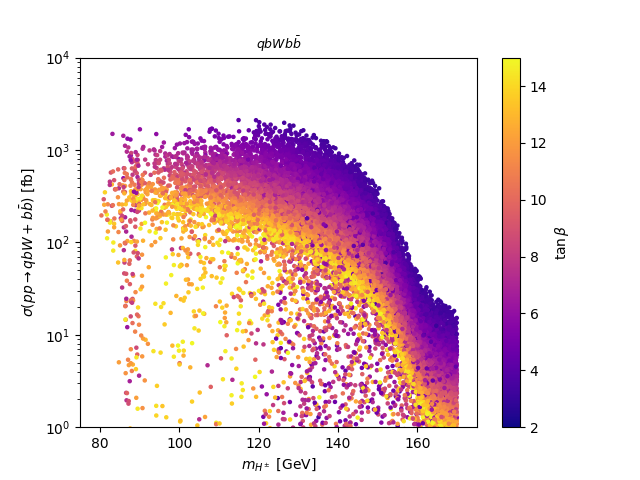}
		\caption{}
	\end{subfigure}
	\hfill
	\begin{subfigure}[t]{.48\textwidth}\centering
		\includegraphics[height=5.8cm,width=7.8cm]{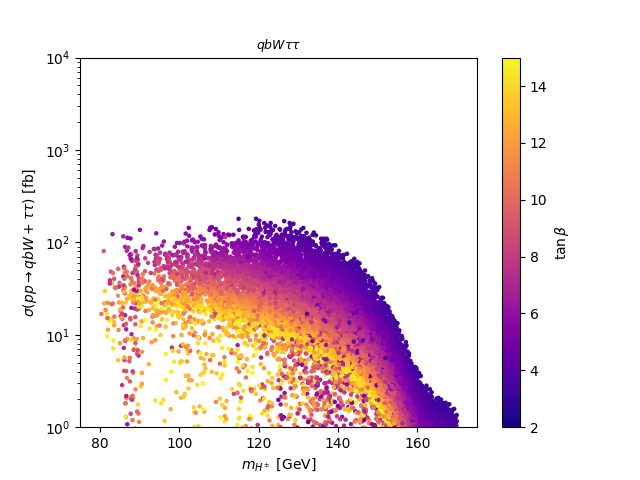}
		\caption{}
	\end{subfigure}
	\\[1cm]
	\begin{subfigure}[t]{.48\textwidth}\centering
		\includegraphics[height=5.8cm,width=7.8cm]{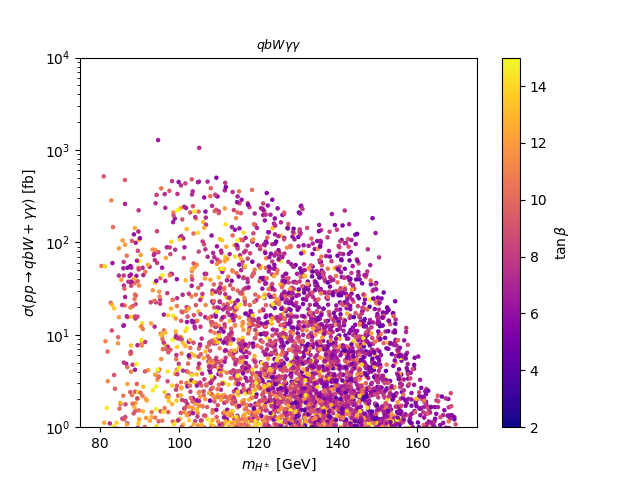}
		\caption{}
	\end{subfigure}
	\caption{\label{fig2} (a) $\sigma(pp\to qbW+\tau\tau)$, (b) $\sigma(pp\to qbW+\tau\tau)$ and (c) $\sigma(pp\to qbW+\gamma\gamma)$ showed against the charged Higgs mass in the $H$ scenario, with the color code indicates the parameter $\tan\beta$.}
\end{figure}

The production rates of the important final states from various scenarios are shown above. In Figure \ref{fig1}, we show the final states; (a) $qbW+b\bar{b}$, (b) $qbW+\tau\tau$ and (c) $qbW+\gamma\gamma$ as a function of charged Higgs mass $m_{H^\pm}$, mapped over $\tan\beta$ in both $h$ and $H$ scenarios. From these figures one can see that, such signatures are interesting for the small $\tan\beta$ (since the coupling of charged Higgs boson to fermions is proportional to $1/\tan\beta$) and the charged Higgs mass in the range between $115$ and $160$ GeV. Similar to Figure \ref{fig1}, we show in Figure \ref{fig2} the same final states in the $H$ scenario. It is clear that one can reach the same conclusion as in $h$ scenario with exception that the final state $qbW+\gamma\gamma$ has a sizable rate in the $H$ scenario. These findings demonstrate that $H^\pm bj$ production channel potentially provide important discovery options for light charged Higgs bosons at the LHC in the 2HDM type I. 

\label{section4}
\section{Conclusion}
In this study,  we have looked for light charged Higgs signatures via the process $pp \to H^\pm b j$ at LHC with $\sqrt{s}=14$ TeV in the 2HDM type I,  taking into account the theoretical and available experimental constraints.  We mainly focused on the bosonic decays of $H^\pm$ and investigated the final states $qbW + b\bar{b}, \tau\tau$ and $\gamma\gamma$ in both $h$ and $H$ scenarios. We have demonstrated that such signatures can be exploited as possible discovery modes for light $H^\pm$ states, even when the signals are contaminated by the QCD background.
\label{section5}

\section*{References}
\bibliography{Refferences.bib}
\end{document}